\documentstyle[prl,multicol,aps,psfig]{revtex} % for camera-ready compuscript
\begin{document}
%
%\twocolumn
\draft
\title{ Numerical Solution of Hard-Core Mixtures }
\author{Arnaud Buhot and Werner Krauth
\cite{email}}
\address{CNRS-Laboratoire de Physique Statistique,
Ecole Normale Sup\'{e}rieure,
24, rue Lhomond,\\ 75231 Paris Cedex 05, France}
\date{Received \today}
\maketitle
\begin{abstract}
We  study the equilibrium phase diagram of binary mixtures of 
hard spheres as well as of parallel hard cubes.
A superior cluster algorithm allows us to establish
and to access the demixed phase for both systems
and to investigate the subtle interplay between short-range depletion
and long-range demixing.
\end{abstract}

\pacs{PACS numbers:  64.75.+g 61.20.Gy} 
\begin{multicols}{2}
\narrowtext 
Liquid binary mixtures pose an important, yet easily formulated
problem:  Imagine a 3-dimensional box of size $L$ (volume $V=L^3$)
occupied by objects of two different types $a, b$ with packing fractions
$\eta_a$ and $\eta_b$.  Will the system  remain homogeneously mixed
or will it phase-separate?

This demixing problem can be readily analyzed whenever there is an
obvious free energy imbalance between the homogeneous and the
phase-separated system.  In many cases of practical importance,
such an imbalance is due to the interactions, for example caused
by electrostatic screening. Even in the absence of interactions
(other than by a hard-core term), a free energy difference can be
caused by an entropic contribution.  In  ``non-additive''
mixtures, the, say, demixed phase may be able to  pack
space more densely. At high overall packing fractions, the system will then
be phase-separated.

Systems of impenetrable large and small spheres or cubes belong to
the class of {\em additive} mixtures. In these systems, the distance
$r_{ab}$ of closest approach between, say, two spheres of radii
$r_a$ and $r_b$ satisfies: $r_{ab} = r_{a} + r_{b}$, so that the
abovementioned simple entropic effect is absent.  Nevertheless, it
has been understood for a long time that even additive mixtures
are subject to an entropic ``depletion force'' \cite{Asakura}:
two {\em large} particles may approach sufficiently closely for
the {\em small} ones to be expelled from the interspace between them. In
that situation, an osmotic (partial) pressure difference (of the
small particles) builds up and pulls the big particles even closer
together. 
The depletion force is strongly attractive at very short distances
$r$ between the large particles (for $2\; r_a < r \lesssim
2\; r_a + r_b$), but finally turns out to be quite long-range in
nature \cite{Mao} \cite{Biben96}.  Approaches to integrate out the
small particles are thus problematic.  As a prototype of additive
mixtures, it is thus of great interest to completely analyze
the microscopic model of large and small spheres or cubes and to
understand whether it will eventually lead to phase separation. 
Precisely this question has remained hotly debated even
in recent years. 

The theoretical framework for the mixture problem has traditionally
relied on the solutions of `closure approximations' to the
Ornstein-Zernicke integral equations.  Classic work of Lebowitz
and Rowlinson \cite{Lebowitz}, performed more than 30 years ago,
first showed that the Percus-Yevick closure of the integral equations
did not lead to phase separation. This exact statement was thought
to reflect the true behavior of the system until Biben and Hansen
\cite{Biben91} challenged the view by  showing that an instability
of the mixture was predicted by a different choice of the closure. Since
the closure approximations are contradictory (and fundamentally
uncontrolled), it is of prime importance to resort to independent
checks.  However, numerical simulations have been notoriously
difficult, especially for
objects very different in size. Numerical evidence for a phase
transition has, to our knowledge, only been obtained in a lattice
model of hard cubes \cite{Dijkstra}.

The situation has thus been very confusing. One is lead to agree
with the author of ref. \cite{Cuesta}: the field would benefit from
an ``Ising model of liquids'', an exactly solvable full-complexity
model against which the concepts and the approximate theories could be
checked. In the present paper, short of providing such an analytic
solution, we obtain a very precise {\em numerical} solution to the
problem of binary mixtures, by applying an extremely powerful
cluster algorithm \cite{Dress} to the problem.  Like the algorithms
of Swendsen and Wang \cite{Swendsen} and Wolff \cite{Wolff} for the
case of  the Ising model, the new method allows to obtain all the
thermodynamic quantities with unprecedented accuracy.
As a first important
application, we actually establish the long sought for demixing
transition {\em both for spheres and for cubes}.

In our algorithm \cite{Dress}, the convergence problems of ordinary
Monte Carlo simulations are completely eliminated in a wide and
physically interesting range of parameters. We have obtained
convergence for up to $10^6$ particles (limited by the size of the
computer memory available to us) where previous work (for systems
two orders of magnitude smaller) remained inconclusive.  Our method
applies equally well to spheres, cubes or any other shape, and both
to the continuum and the lattice \cite{Heringa}.

At any step of the algorithm (cf. \cite{Dress} for  details) one
generates a new `copy' of the `original' by inverting the latter
around a randomly chosen pivot point $x_p$.  Original and copy are
then superimposed. The combined system presents `clusters' of
overlapping objects. Some of these clusters are then `flipped':
particles belonging to the original are assigned to the copy, and
vice versa.  Thereafter, the copy is discarded and a new pivot
point is chosen.  The algorithm can easily be set up for simulations
at constant particle number. It is completely symmetric with respect
to the operation:  original $\rightleftharpoons$ copy, a property
which implies detailed balance. The method works perfectly well as
long as the combined system breaks up into at least a few sizeable
clusters. In our previous work on hard spheres in two dimension,
we located this purely algorithmic percolation threshold at a much
lower packing fraction than would have been useful to study the 2-dimensional
liquid-solid phase transition. In the present case of mixtures,
the situation is vastly improved: we find
that the percolation threshold (the optimal operation point of our
algorithm)  mainly depends on the {\em combined packing fraction} $\eta_a + 
\eta_b$, but {\em not on the size ratio} of the two species.
This is radically different from the behavior of ordinary (local)
Monte Carlo methods, where the diffusion of large particles becomes
completely blocked by the nearby presence of many small ones.  The
lack of sensitivity of our algorithm to the size ratio of the
particles is all the more interesting for a second reason: it has
long been understood that the important physical phenomena
(depletion and the tendency to phase-separate) quickly move to
lower overall  packing fractions as the particles become dissimilar in size.
These two effects open up a large window of packing fractions and size
ratios in which we can numerically solve the problem of liquid
mixtures with the new algorithm.

Let us first consider the superposed system of {\it monodisperse}
objects. In  this case, the algorithm's optimal point of operation
turns out to be at $\eta_a = \eta_P \sim 0.23$ both for cubes and
for spheres. The percolation threshold is thus located at a packing fraction 
corresponding to $1/4$ of the close packing fraction for cubes and
$1/3$ for spheres.  At $\eta_P$, the algorithm evenly flips clusters
of any size. For larger packing fractions, intermediate cluster sizes will
appear less often, since we either encounter the percolating cluster,
or have to do with the algebraically decreasing distribution of
small clusters \cite{Stauffer}.  Above the threshold, the algorithm
deteriorates `gracefully', and it is quite possible to converge
the monodisperse system (at, say, $N=500$) for packing fractions up to
$\eta \sim 0.4$.

We now introduce the very large number of  small objects (cubes or
spheres). As long as the system  remains homogeneous, we notice
that the percolation threshold  moves to slightly larger {\it
overall} packing fractions.  During the simulation, we  sample the
partial distribution function of pairs of large particles $g_{ll}(r)$
\cite{Hansen}. The numerical noise is much reduced if we consider
not $g_{ll}(r)$, but the {\em integrated} pair distribution function
$G_{ll}(r) = 4 \pi \rho_l \int^r dr' r'^2 g_{ll}(r')$ ($\rho_l$
is the density of large particules).  $G_{ll}(r)$  determines the average
number of large particles within a radius $r$ around a randomly
chosen large particle.  We also compare $g_{ll}(r)$ and $G_{ll}(r)$
to the pair distribution functions $g(r)$ and $G(r)$ of the
monodisperse system (with $\eta_b=0$) at the same value of $\eta_a$.
Knowledge of the pair distribution functions for all $r$ is equivalent
to the computation of the structure factor (its Fourier transform),
which informs us about the system's phase:  for a homogeneous phase,
$g_{ll}(r)$ is completely flat for large arguments ($G_{ll}(r)$
will follow the monodisperse system's $G(r)$).  In contrast,
$g_{ll}(r)$ and $G_{ll}(r)$ for  phase separated systems will be
system-size dependent functions even for moderate $r$. This fact
translates to the presence of phase regions of varying extension.
\begin{figure} 
\centerline{ \psfig{figure=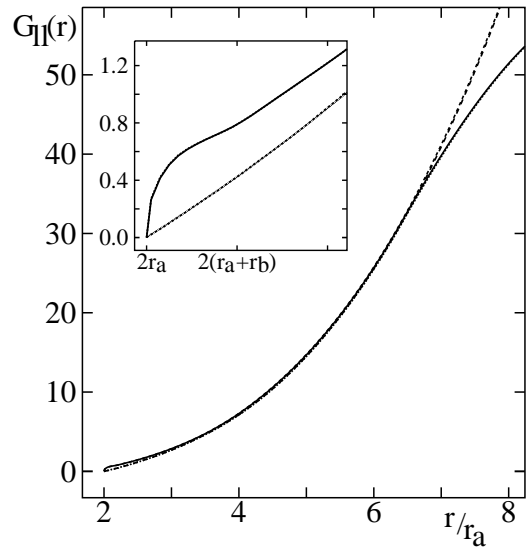,height=8cm}}
\caption{ Integrated pair distribution function $G_{ll}(r)$ for
spheres of radii $r_b/r_a = 1/10$ ($N_a = 62, N_b= 62.000$ and
$N_a= 108, N_b=108.000$) (upper), compared to the monodisperse case
(lower). The packing fraction is $\eta_a = \eta_b = 0.1215$, the
system is homogeneous.  } \end{figure}

For spheres  at a packing fraction $\eta_a = \eta_b = 0.1215$, and a ratio
of radii $r_b/r_a = 1/10$, we determined the integrated pair
distribution function $G_{ll}(r)$ and compared it to the  monodisperse
system ($\eta_b=0$). In fig. 1, we present our results for $N_a =
62, N_b= 62.000$ and for $N_a= 108, N_b=108.000$ (the latter
case, {\em e. g.}, corresponds to a box of side length $L=15.5 \times
r_a$, with periodic boundary conditions).  We obtain very
smooth curves, which indicate the exceptional convergence of the
algorithm.  More importantly, close agreement of $G_{ll}(r)$ with
the monodisperse case for large $r$  is reached. This clearly
indicates that the introduction of small spheres has not changed
the large-scale behavior of our system, which is still homogeneous.
The inset of fig. 1 shows the $G_{ll}(r)$ for small arguments: the
lower line corresponds to the monodisperse system, and the upper
curve to the full simulation. A dramatic depletion effect is obvious.
In agreement with previous knowledge, we observe that the effect
is strongest for values $ 2 r_a < r \lesssim  2 (r_a +
r_b)$.  Differentiating $G_{ll}(r)$, we obtain $g_{ll}(r)$, whose
contact value  at $r= 2 r_a$ is increased by a factor of $8.5$ with
respect to the monodisperse case.  $g_{ll}(r)$ oscillates for larger
arguments, reaches a first minimum at $r \sim 2 (r_a + r_b)$, and
eventually levels out to the expected value $g_{ll}(r) \rightarrow 1$
for large $r$.
For small $r$, the integrated function, $G_{ll}(r)$,
exceeds the monodisperse system's $G(r)$ by about $0.4$. In our
opinion, this additional binding of an average $0.4$ particles
characterizes the strength of the depletion much better than the
contact value $g_{ll}(2 r_a)$. We also suggest that a system with
an additional binding of more than one particle should be unstable
to phase separation.   

The inset of fig. 1, as the main graph, contains in fact two sets
of curves. On the scale of the figure, the curves for $N_a = 62$
cannot be distinguished from those at $N_a = 108$ (for $r<L/2$,
neither can the data for $N_a = 864$, which we have also computed).
As mentioned before, the close agreement testifies to the good
convergence of the algorithm, but also indicates that finite-size
effects are completely negligible at these values of the  physical
parameters.  This is a key observation, which leads us to strongly
suspect that we are far away from any second-order phase transition
point (as the critical point of phase separation), which should
lead to such effects. 

Finally, at the combined packing fraction of
$\eta=0.243$, we are still {\em below} the percolation threshold
for the system with $r_b/r_a = 1/10$, but clearly {\em above} the
monodisperse system's point of percolation.  We believe that the
local binding effects of the depletion force lead to a slight
modification of the many-particle distribution functions, which is
picked up in the distribution of cluster sizes.

The situation appears to be changed for a ratio of radii $r_b/r_a=
1/20$.  For this case, our memory resources have allowed us to
perform calculations at $N_a = 32, N_b= 256.000$ and $N_a = 62,
N_b= 496.000$, again at $\eta_a = \eta_b = 0.1215$ ($L=10.33
\times r_a$ and $L=12.88 \times r_a$, respectively). Phase
separation still does not seem to have taken place, but we notice
the presence of important 
finite-size effects. For example, the system at $N_a = 32$ indicates
an average `additional binding' of $0.7$ particles per sphere, while 
the larger system at $N_a = 62$ yields a binding of $0.8$.
Again, the finite-size effects go into the direction of an
{\em increased depletion} for the larger size. Since we are already
at the limit of our computer resources, we were unable to check
the phase behavior at even larger numbers of particles.

Finally, we have performed simulations at $r_b/r_a= 1/30$, and
$N_a = 32, N_b= 864.000$, again at the same packing fraction $\eta_a
= \eta_b = 0.1215$. In these very large simulations, the system
clearly has crossed into the separated phase:  the distribution
function $G_{ll}(r)$ is suddenly very different from the monodisperse 
system's $G(r)$ for all values of $r$, and the additional binding is 
clearly larger than $1$. In this case, we usually observe the presence of
one or two large aggregates \cite{footaggregate} 
of large spheres which comprise most of them.  In agreement with this
observation we also notice a dramatic change in the behavior of
our algorithm:  the distribution of `cluster' sizes is 
shifted towards very large clusters, since the presence of the
dense phase of spheres (the `aggregate') pushes the system locally
beyond the percolation threshold.

Our algorithm is extremely powerful, but we have nevertheless
reached the memory limits of today's  workstations.  For spheres,
the transition takes place at rather small size ratios ($r_b/r_a
\sim 1/20$):  therefore, we simulate a very large total number of
particles but the $G_{ll}(r)$ belongs to a small system with only
a few dozen large spheres. For spheres, the  finite-size analysis,
and the precise location of the critical point will
have to be done on more powerful machines.

\begin{figure}
\centerline{ \psfig{figure=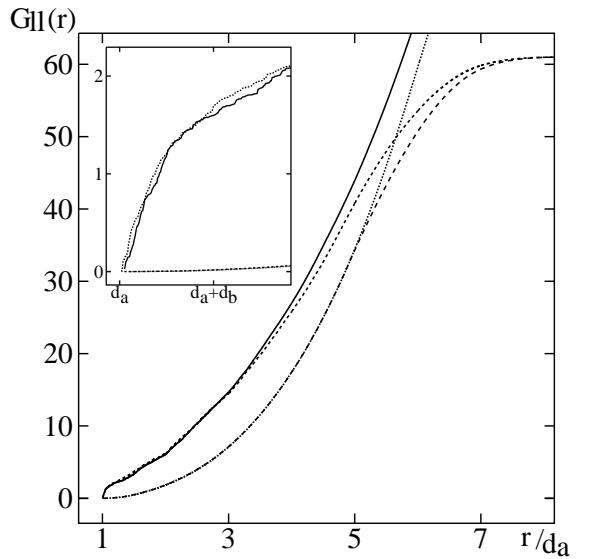,height=8cm} }
\caption{ $G_{ll}(r)$ for cubes of side lengths  $d_b/d_a = 1/10$
(upper two curves at small $r$), compared to the monodisperse
case (lower) for $N_a = 62, N_b= 62.000$ and $N_a = 108, N_b=
108.000$. The system, at a packing fraction of $\eta_a = \eta_b =
0.067$ is phase-separated.  } 
\end{figure}

We have found it interesting to pursue our study in a different
direction, {\em i. e.} to confirm phase separation in the model of
hard parallel cubes of side lengths $d_a$ and $d_b$, respectively.
Here, the large cubes can touch on opposite faces.  The osmotic
pressure, exerted on a much larger surface, leads to a stronger
depletion force, which should strongly favor phase separation.
This is indeed what we have found.  As for hard spheres, we have
observed {\it i)} depletion at any packing fraction; {\it ii)} finite-size
effects, which become more important as the cubes grow more dissimilar
in size, and which render the system more unstable for larger
particle numbers;  {\it iii)} the phase-separated regime for very
strong depletion.  Again, the instability appears as soon
as the average additional `binding' is of one particle per cube.

Simulations have revealed no signatures of an instability for a
size ratio of $d_b/d_a=1/2$ (at the accessible packing fractions
below $\eta_a + \eta_b = \eta_P \sim 0.23$). Finite-size effects
are negligible.  For a ratio of $d_b/d_a=1/10$, at small packing
fractions, the same holds true.  However, for a packing fraction
of $\eta_a = \eta_b = 0.054$, the system at $N_a = 62 ; L=10.5
\times d_a$ remains clearly homogeneous, with a very strong depletion
and an `additional binding' of about $0.9$ particles. At the same
packing fraction, at $N_a = 496, N_b=496.000 ; L=21.0 \times d_a$,
the $G_{ll}(r)$ has pulled away from the monodisperse case for all
$r$: the system has already undergone phase separation.  In fig.
2, we present well-converged data for a slightly larger packing
fraction  $\eta_a = \eta_b = 0.067 $, still at $d_b/d_a=1/10$.
Here, already the system at $N_a = 62 ; L=9.75 \times d_a$ is in
the demixed phase.  The data at $N_a=108 ; L=11.73 \times d_a$ 
illustrate the large
finite-size effects at moderate $r_a$. The maximum difference of $(G_{ll}
- G)$ is much larger for $N_a=108$ than for $N_a=62$, and is expected
to diverge for $N_a \rightarrow +\infty$. We stress again that 
this effects are specific of the phase-separated system,
and in the present form of the fixed particule number system, in which
actual phase coexistence is obtained.
In the demixed system, one of the two phases will be
`rich in cubes'. Unfortunately this phase will have a packing
fraction above $\eta_P$ (except very close to the critical point),
and will be difficult to study with our algorithm.  The route
towards instability is {\em not} touched by this problem.
Nevertheless, both systems seem to have
converged. While we did not study the demixed phases in detail,
we have noticed the appearance of aggregates which are all but
closed packed.
 
In conclusion, we have studied the equilibrium phases of hard core
mixtures. A superior algorithm has allowed us to establish and to
access the demixed phase both for spheres and for cubes, and to
investigate the subtle interplay between short-range depletion
and long-range demixing.  There are  many questions and a large
number of directions for further research, besides those already
mentioned.  Primarily, we think that the precise phase diagram
needs to be established, especially the position of the critical
point. In addition, the comparison with various closure formulas
should be undertaken. We can already see that
ref. \cite{Cuesta} places the critical packing fraction 
for cubes much too high.
For most closure approximations,
the numerical applications were done at quite large values of the
size ratio, probably since Monte Carlo simulations for very dissimilar
objects were thought to be completely out of reach.  Paradoxically,
the opposite is true. We are much more at ease at large asymmetry,
as long as the packing fraction is not too high.  
The comparison between the exact numerical points and the closures
should of course be done directly on the observables, such as
$g_{ll}(r)$, and not on the phase diagram. To encourage and simplify
further work, we will make available via email the Fortran code
used in this paper.

Finally, the question remains whether the artificial percolation
threshold $\eta_P$ presents an unsurmountable barrier to the numerical
solution.
Several  ideas to go  much beyond $\eta_P$ have been
formulated (cf  \cite{Dress}). Even in the very dense limit of
importance in the two-dimensional melting problem, the flip of the
percolating cluster can be avoided, but we have up to now been
unable to transform this idea into a working algorithm \cite{Bagnier}.
However, we are firmly convinced that $\eta_P$ is not a hard
boundary: many more difficult problems, such as melting, are also
likely to fall under the attack of appropriate, while very
specialized algorithms.

\end{multicols}

\begin{references}
\bibitem[*]{email} buhot@physique.ens.fr; krauth@physique.ens.fr

\bibitem{Asakura} S. Asakura and F. Oosawa
{\sl J. Chem. Phys.} {\bf 22} 1255 (1954).

\bibitem{Mao} Y. Mao, M. E. Cates and H. N. W. Lekkerkerker
{\sl Physica A} {\bf 222} 10 (1995).

\bibitem{Biben96} T. Biben, P. Bladon and D. Frenkel
{\sl J. Phys. Condens Matter}  {\bf 8} 10799 (1996).

\bibitem{Lebowitz} J. L. Lebowitz and J. S. Rowlinson
{\sl J. Chem. Phys.} {\bf 41}, 133 (1964). 

\bibitem{Biben91} T. Biben and J. P. Hansen 
{\sl Phys. Rev. Lett.} {\bf 66}, 2215 (1991).

\bibitem{Dijkstra} M. Dijkstra, D. Frenkel and J. P. Hansen
{\sl J. Chem. Phys.} {\bf 101}, 3179 (1994); M. Dijkstra and D.
Frenkel {\sl Phys. Rev. Lett.} {\bf 72} 298 (1994).

\bibitem{Cuesta} J. A. Cuesta {\sl Phys. Rev. Lett.} {\bf 76} 3742
(1996).

\bibitem{Dress}
C. Dress and W. Krauth {\sl J. Phys. A: Math Gen} {\bf 28}, L597 (1995).

\bibitem{Swendsen} 
R. H. Swendsen and J.-S. Wang {\sl Phys. Rev.  Lett.} {\bf 58} 86
(1987).

\bibitem{Wolff} U. Wolff {\sl Phys. Rev. Lett.} {\bf 62} 361 (1989).

\bibitem{Heringa} 
The first successful simulations using the algorithm were performed
in a $3$-dimensional lattice gas model:
J. R. Heringa and H. W. J. Bl\"{o}te 
{\sl Physica} {\bf 232A}, 369 (1996); {\it preprint}
 to appear in Physica A; J. R. Heringa {\it preprint}.

\bibitem{Stauffer}
D. Stauffer {\sl Phys. Rep. } {\bf 54} 1 (1979).

\bibitem{Hansen} J. P. Hansen and I. R.  Macdonald {\sl Theory of
Simple Liquids, 2nd edition} (Academic, London, 1986).

\bibitem{footaggregate} Two large spheres are `aggregated' 
if their distance is smaller than $ 2( r_a + r_b)$.

\bibitem{Bagnier} S. Bagnier and W. Krauth, unpublished (1996).

\bibitem{footcheck} The algorithm was carefully checked against 
a local Monte Carlo method for small systems. In one dimension, 
excellent agreement with the exact solution \cite{Robledo} was obtained.

\bibitem{Robledo} A. Robledo and J. S. Rowlinson
{\sl Molec. Phys. } {\bf 58}, 711 (1986).
\end{references}
\end{document}